\documentclass{article}

\usepackage{arxiv}

\usepackage[utf8]{inputenc} 
\usepackage[T1]{fontenc}    
\usepackage{hyperref}       
\usepackage{url}            
\usepackage{booktabs}       
\usepackage{amsfonts}       
\usepackage{nicefrac}       
\usepackage{microtype}      
\usepackage{amsmath}
\usepackage{cleveref}       
\usepackage{lipsum}         
\usepackage{graphicx}
\usepackage{natbib}
\usepackage{doi}
\usepackage{xcolor}
\usepackage[colorinlistoftodos,prependcaption]{todonotes}
\usepackage{pdfpages}

\usepackage{subcaption}

\usepackage{booktabs}
\usepackage{threeparttable}
\usepackage{array}
\usepackage{ragged2e}
\usepackage{makecell}
\usepackage{tabularx}
\newcolumntype{Y}{>{\RaggedRight\arraybackslash}X}

\usepackage[version=4]{mhchem}

\title{Investigation of regional variations in CO$_2$ growth rates : Integrating Emission Inventories and Atmospheric Observations }

\newif\ifuniqueAffiliation
\uniqueAffiliationtrue

\ifuniqueAffiliation 
\author{ \href{https://orcid.org/0009-0003-2699-5296}{\includegraphics[scale=0.06]{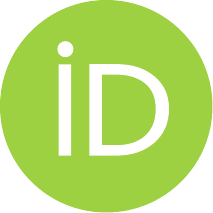}\hspace{1mm}Yogesh Bali}\\
	Institut f\"ur Mathematik \\
	 Johannes Gutenberg-Universit\"at Mainz\\
	Staudingerweg 9, 55099 Mainz, Germany \\
	\texttt{ybali@uni-mainz.de} \\
	\And
	\href{https://orcid.org/0000-0002-3649-719X}{\includegraphics[scale=0.06]{orcid.pdf}\hspace{1mm}Darja Cvetković} \\
	Scientific Computing Labaratory\\
	Institute of Physics Belgrade\\
	Belgrade, Serbia \\
	\texttt{darja@ipb.ac.rs} \\
    \And
	\href{https://orcid.org/0000-0002-0673-174X}{\includegraphics[scale=0.06]{orcid.pdf}\hspace{1mm}Juan Gancio} \\
	Departamento de Física\\
	Universitat Politècnica de Catalunya\\
	Terrassa, Spain \\
	\texttt{juan.gancio@upc.edu} \\
    \And
	\href{https://orcid.org/0009-0006-0214-3701}{\includegraphics[scale=0.06]{orcid.pdf}\hspace{1mm}Adrián Gutiérrez-Arroyo} \\
	Grupo Interdisciplinar de Sistemas Complejos\\
	Universidad Carlos III de Madrid\\
	Leganés, Spain \\
	\texttt{adgutier@math.uc3m.es} \\
    \And
	\href{https://orcid.org/0000-0002-1541-8683}{\includegraphics[scale=0.06]{orcid.pdf}\hspace{1mm}Sofia {Vazquez Alferez}} \\
	Department of ICS, Department of Mathematics\\
	Utrecht University\\
	Utrecht, The Netherlands \\
	\texttt{s.vazquezalferez@uu.nl} \\
    \And
    \href{https://orcid.org/0009-0009-6708-9517}{\includegraphics[scale=0.06]{orcid.pdf}\hspace{1mm}Xuan Tung Vu} \\
	Laboratoire d'Ingénierie des Systèmes Complexes\\
	INRAE, Université Clermont Auvergne\\
	Clermont-Ferrand, France \\
	\texttt{xuan-tung.vu@inrae.fr} \\
    \And
	\href{https://orcid.org/0009-0007-9518-0161}{\includegraphics[scale=0.06]{orcid.pdf}\hspace{1mm}Jin Yan} \\
	Weierstrass Institute for Applied Analysis and Stochastics\\
	Berlin, Germany\\
	\texttt{yan@wias-berlin.de} \\
    \And
	\href{https://orcid.org/0009-0007-4083-1078}{\includegraphics[scale=0.06]{orcid.pdf}\hspace{1mm}Pietro Zgaga} \\
	Networks Unit\\    
    IMT School of Higher Studies Lucca \\
    Lucca, Italy\\
	\texttt{piezga@protonmail.com} \\
    \And
	\href{https://orcid.org/0000-0002-2511-7673}{\includegraphics[scale=0.06]{orcid.pdf}\hspace{1mm}Fakhteh Ghanbarnejad} \\
    School of Technology and Architecture\\
	SRH University of Applied Sciences Heidelberg\\
	Leipzig Campus, Germany\\
	\texttt{fakhteh.ghanbarnejad@gmail.com} \\
    \And
	\href{https://orcid.org/0000-0003-1295-6555}{\includegraphics[scale=0.06]{orcid.pdf}\hspace{1mm}Nasrin Mostafavi Pak} \\
	Institute of Envionmental Physics\\
	University of Bremen\\
	Bremen, Germany\\
	\texttt{mostafav@uni-bremen.de} \\
}

\hypersetup{
pdftitle={A template for the arxiv style},
pdfsubject={q-bio.NC, q-bio.QM},
pdfauthor={David S.~Hippocampus, Elias D.~Striatum},
pdfkeywords={First keyword, Second keyword, More},
}

\begin{document}
\maketitle

\begin{abstract}

Atmospheric carbon dioxide (CO$_2$) growth reflects the combined influence of anthropogenic emissions, biospheric carbon exchange, and climate variability. However, climate mitigation efforts are primarily evaluated using bottom-up emission inventories reported within political boundaries, creating a fundamental disconnect between emission accounting and the spatially continuous atmospheric response. Here, we present a global, top-down analysis of atmospheric CO$_2$ growth rates using gridded CAMS atmospheric column CO$_2$ reanalysis product, anthropogenic emissions from EDGAR, biospheric activity derived from GOSIF dataset, and large-scale climate variability represented by the Southern Oscillation Index (SOI). We investigate how regional atmospheric CO$_2$ growth responds to both human and natural drivers and identify the dominant controls across the globe.

Our analysis reveals that atmospheric CO$_2$ growth rates vary significantly in space and time, but these variations are predominantly driven by natural carbon-cycle processes and global background trends. The signal from anthropogenic emissions is frequently masked by this natural variability, making the top-down detection of emission changes challenging in many regions. The year 2020 offers a unique natural experiment: despite substantial declines in fossil fuel emissions reported in bottom-up inventories due to COVID-19 lockdowns, and despite being a neutral ENSO year (so no strong climate-driven signal), the resulting reductions were not consistently reflected in local atmospheric CO$_2$ growth rates. This spatial decoupling underscores that even in the absence of strong climate forcing, local atmospheric responses are dominated by biospheric dynamics, atmospheric transport, and regional carbon-cycle memory.

To better understand these regional differences, we classify the global land surface into characteristic carbon-cycle regimes using unsupervised clustering and persistence analysis. We identify five dominant regimes—ranging from carbon-inactive polar regions to anthropogenic cores and active biospheric sinks. However, we find that spatial averaging within these clusters largely smooths out unique regional growth patterns, leaving the global climate signal as the primary driver in most regimes. A notable exception is the active biosphere cluster, where the biogenic signal remains strong enough to survive the smoothing process, highlighting the unique and dominant role of tropical forests in the global carbon cycle.
\end{abstract}


\section{Introduction}\label{sec:Intro}

Climate change is one of the greatest challenges that  humanity faces in the current century \citep{IPCC2023WGII}. Not only it involves devastating consequences for humanity itself, as the reduction of food production, increase of poverty and inequality, uninhabitability of populates areas, and  increase in the frequency and severity of extreme events, like floods, heatwaves, and droughts, but also the destruction of our planet as we know it: reduction of the polar ice-caps and glaciers, increase sea level, species extinction, deforestation, coral bleaching, etc \citep{IPCC2023WGII,IPCC2019SROCC}.  As a complex system, the drivers behind the global temperatures form an intricate network of interactions and feedback loops that have allowed relatively stable temperatures throughout the Holocene.

Natural external forcings include variations in incoming solar radiation\citep{Solanki2013SolarIrradiance}, volcanic aerosol forcing\citep{Cole-Dai2010VolcanoesClimate}, and long-term orbital forcing\citep{Lourens2021ChapterChange}. In addition, internal ocean--atmosphere variability such as the El Ni\~no--Southern Oscillation (ENSO), an irregular warming and cooling of the tropical Pacific Ocean , contributes to interannual fluctuations in global and regional climate \citep{Yeh2009ElNino}.

While these natural forcing and internal modes of variability influence climate, the recent long term warming trend is primarily associated with anthropogenic driven changes in the atmospheric conditions, espically the increase in greenhouse gasses (GHG) concentrations \citep{IPCC2023WGII,IPCC2019SROCC}.
A central mechanism is the atmospheric greenhouse effect, which depends on the absorption and re-emission of terrestrial infrared radiation by greenhouse gases
such as water vapour, carbon dioxide (CO$_2$), methane (\ce{CH4}), and nitrous oxide (\ce{N2O}) \citep{Arrhenius1896XXXI.Ground,Tyndall1861XXIII.Lecture}.

Up until the industrial revolution (approximately 1760–1840), atmospheric GHG concentrations, particularly CO$_2$ was shaped by the balance between natural carbon sources and sinks, including photosynthesis, respiration, ocean exchange, and long-term carbon-cycle processes \citep{Isson2020GlobalCarbonCycle}. Since then,
human activities have strongly perturbed this balance through fossil-fuel combustion, cement production, agriculture, and land-use change. Anthropogenic
CO$_2$ emissions arise mainly from fossil-fuel combustion, cement production, and land-use change, with land-use-change emissions closely linked to
deforestation \citep{friedlingstein2026global}. These anthropogenic changes are now the dominant driver of the recent increase in global mean surface temperature \citep{IPCC2023WGII,IPCC2019SROCC}.

In 2015, 195 countries signed the Paris Agreement, which aimed to maintain global warming below $2^{\circ}$C above pre-industrial levels, while at the same time attempting to limit this temperature to $1.5^{\circ}$C \citep{Delbeke2019ClimateNeutralEurope}. 
At the time the agreement was signed, record breaking temperatures were registered, as a consequence of a particularly strong El Niño event, reaching a global temperature average of $1^{\circ}$ above pre-industrial levels \citep{2016AYear,Tollefson20162015Record,Mann2016TheWarmth}. The record was broken again the following year \cite{Kennedy2017Global2016}, and again in two consecutive years: 2023 and 2024, the last one exceeding the barrier of $1.5^{\circ}$C \citep{Forster2024IndicatorsGlobalClimateChange2023,Forster2025IndicatorsGlobalClimateChange2024}. 2025 became the third warmest year on record (after 2024 and 2023), when surface temperatures remained high despite a neutral phase of ENSO. The period 2023--2025 represents the first three year period with an average above $1.5^{\circ}$C pre-industrial levels, and all the years in the period 2015--2025 are the warmest 11 years on record \citep{C3S2026GlobalClimateHighlights2025}.

According to the Intergovernmental Panel on Climate Change (IPCC), in order to achieve the long term goals of the Paris agreement, GHG emissions must be reduced by 43\% by 2030, with respect to 2019 levels \citep{IPCC2023ClimateChange}. Given their production of GHG, their vulnerability to climate change, and their developmental needs, different goals were assigned to different countries in the agreement \citep{Pauw2019SubtleAgreement}.
Although GHG emissions have constantly increased since the industrial revolution \citep{Crippa2025GHGEmissionsAllWorldCountries}, their production has been heterogeneous across countries, with several of them achieving a reduction during the last decades \citep{Lamb2022CountriesSector}.  However, tracking the actual progress in meeting the GHG emission reduction objectives is based on the aggregation process in which each country reports their results and adaptation measures \citep{Craft2018MeasuringAgreement}. This bottom-up process suffers from many conceptual and methodological problems \citep{Lesnikowski2017AssessingAdaptationProgress}, such as  the use of different conventions, definitions, and offsets, which has led to significant gaps between reported GHG emissions and scientific models \citep{Grassi2021CriticalProgress,Grassi2025ImprovingAgreement,Gidden2023AligningBenchmarks,Allen2025GeologicalSinks}.

Most greenhouse-gas emissions estimates used in climate-policy assessments are based on bottom-up inventories, in which emissions are reconstructed from activity data, sectoral information, and emission factors and are often interpreted within national or
administrative boundaries\citep{IPCC2006GuidelinesGHGInventories,Crippa2024InsightsV8.0}. However, atmospheric CO$_2$ growth is spatially continuous and reflects the combined effects of anthropogenic emissions, biospheric exchange, carbon sinks, and atmospheric transport \citep{friedlingstein2026global}. This creates a
gap between policy-based emission accounting and the spatial structure of observed atmospheric CO$_2$ variability.

In addition, GHG fluxes are  driven not only by human-induced effects but also by natural effects \citep{friedlingstein2026global}, such as the ENSO cycle \citep{Chatterjee2017InfluenceElNinoOCO2}. Because disentangling the contributions of natural and anthropogenic drivers can be challenging \citep{Lamb2026GHGEmissionsEstimatesExplained}, the natural variability of the climate could mask the actual effects of human GHG emissions.

Previous studies have examined related components of atmospheric CO$_2$ variability. Satellite- and ground-based column observations have been used to estimate annual \ce{XCO2} growth rates at global, latitude-band, and station scales, and to examine their variability in relation to surface records, fossil-fuel emissions, ENSO, and short-term emission-reduction events
\citep{Buchwitz2018,mostafavi2026annual}. Other work has used satellite \ce{XCO2} observations to detect anthropogenic emission regions,
for example over the eastern USA, central Europe, and East Asia, or to evaluate bottom-up emission inventories such as EDGAR, particularly over China and other major emitting regions
\citep{hakkarainen2016direct,zhang2022evaluating}. In parallel, atmospheric CO$_2$ observations and solar-induced chlorophyll
fluorescence have been used to study biospheric carbon uptake and event-driven carbon-cycle responses over North America and the tropics \citep{shiga2018atmospheric,liu2017contrasting}.

These studies have generally focused on a particular component: atmospheric growth-rate estimation, anthropogenic emission detection or inventory evaluation, or biospheric carbon-cycle analysis. \citet{mostafavi2026annual} showed that CAMS satellite reanalysis  dataset shows acceptable agreement with ground based total column measurements by TCCON specifically at locations that ground based data are missing or not possible. Using the reanalysis data with global coverage we assess how CO$_2$ growth rates change in different regions of the glob where measurements are missing. 
In this study, we combine gridded
satellite reanalysis dataset from 
CAMS \citep{AgustiPanareda2023CAMS}to calculate atmospheric CO$_2$ growth rates all over the globe, along with the EDGAR anthropogenic emissions \citet{Crippa2025GHGEmissionsAllWorldCountries}, GOSIF
biospheric activity \citet{Li2019AData}, and SOI indices \citet{Ropelewski1987SOI}  %
to provide a global analysis of the
spatial patterns of atmospheric CO$_2$ growth 
and their relationship %
with both emission-related and biosphere-related drivers. 

Recent data-driven studies have used machine-learning methods to estimate gridded or regional anthropogenic \ce{CO2} emissions from satellite-derived \(X\ce{CO2}\), vegetation indicators, night-time lights, and other auxiliary variables \citep{Zhang2022AnthropogenicCO2RandomForest,Ji2024AnthropogenicCO2Clustering,Mustafa2021NeuralNetworkCO2Emissions}. Other work has used machine-learning models to reconstruct high-resolution \(X\ce{CO2}\) fields from satellite, model, and geographic predictors \citep{He2023XCO2MachineLearningChina}. These studies mainly focus on prediction or reconstruction, whereas our study uses an interpretable grid-cell-wise regression framework to examine associations between \ce{CO2} growth rate, anthropogenic emissions, and biospheric activity.

This approach offers a top-down, data driven, complex system approach, which allows to study the atmospheric response on climate change mitigation measures in different regions of the globe. Specifically 2020, as a case study, offers a valuable example of how much abrupt reductions of human induced CO$_2$ emissions may be observed in different regions of the world. As natural variability always plays a role in CO$_2$ growth rate, but in different amount in different regions of the world, this study gives us the opportunity to assess which regions the amount of change in emissions along with other atmospheric conditions allowed the detection of anthropogenic emission changes using measurements. 

In addition, we classify the different regions of the globe according to the two main drivers on CO$_2$ growth rates, which can be linked to specific Earth systems and are related to specific stages of the carbon cycle. 
 
The remainder of this paper is organized as follows. Section \ref{sec:Data} describes the datasets used in the analysis, including CAMS-derived atmospheric CO$_2$ growth rates, EDGAR anthropogenic emissions, GOSIF-derived biospheric activity, and the lagged Southern Oscillation Index. Section \ref{sec:Methods} presents the methodological framework, including pixel-wise regression, unsupervised clustering, persistence analysis, and regression on aggregated carbon-cycle regimes. Section \ref{sec:Results} reports the main results, first examining spatial patterns of observed and predicted CO$_2$ growth rates, then identifying persistent carbon-cycle regimes, and finally analyzing the temporal behavior of these regimes using aggregated regression models. Section \ref{sec:Conclusion} discusses the main implications, limitations, and possible extensions of the framework.

we further identify characteristic carbon-cycle regimes using unsupervised machine-learning techniques, in order to understand and predict regional responses better under different conditions and changes in the drivers. 

\section{Data}\label{sec:Data}
In this work, we used four main datasets: the total column Atmospheric carbon dioxide mole fractions (XCO$_2$) provided by the Copernicus Atmosphere Monitoring Service (CAMS) greenhouse gas reanalysis, anthropogenic CO$_2$ emissions by the Emissions Database for Global Atmospheric Research (EDGARD). The biospheric component of the analysis was represented GOSIF dataset. GOSIF provides a global measure of vegetation photosynthetic activity derived from solar-induced chlorophyll fluorescence (SIF), using a data-driven approach that combines OCO-2 SIF soundings, MODIS remote sensing data, and meteorological reanalysis data. In another approach, the Southern Oscillation Index (SOI), which relates to the state of the ENSO through variation of sea level pressure was used as the biospheric driver. For an easier notation, we will refer to these datasets as:

\begin{figure}
    \centering
    \includegraphics[width=0.9\linewidth]{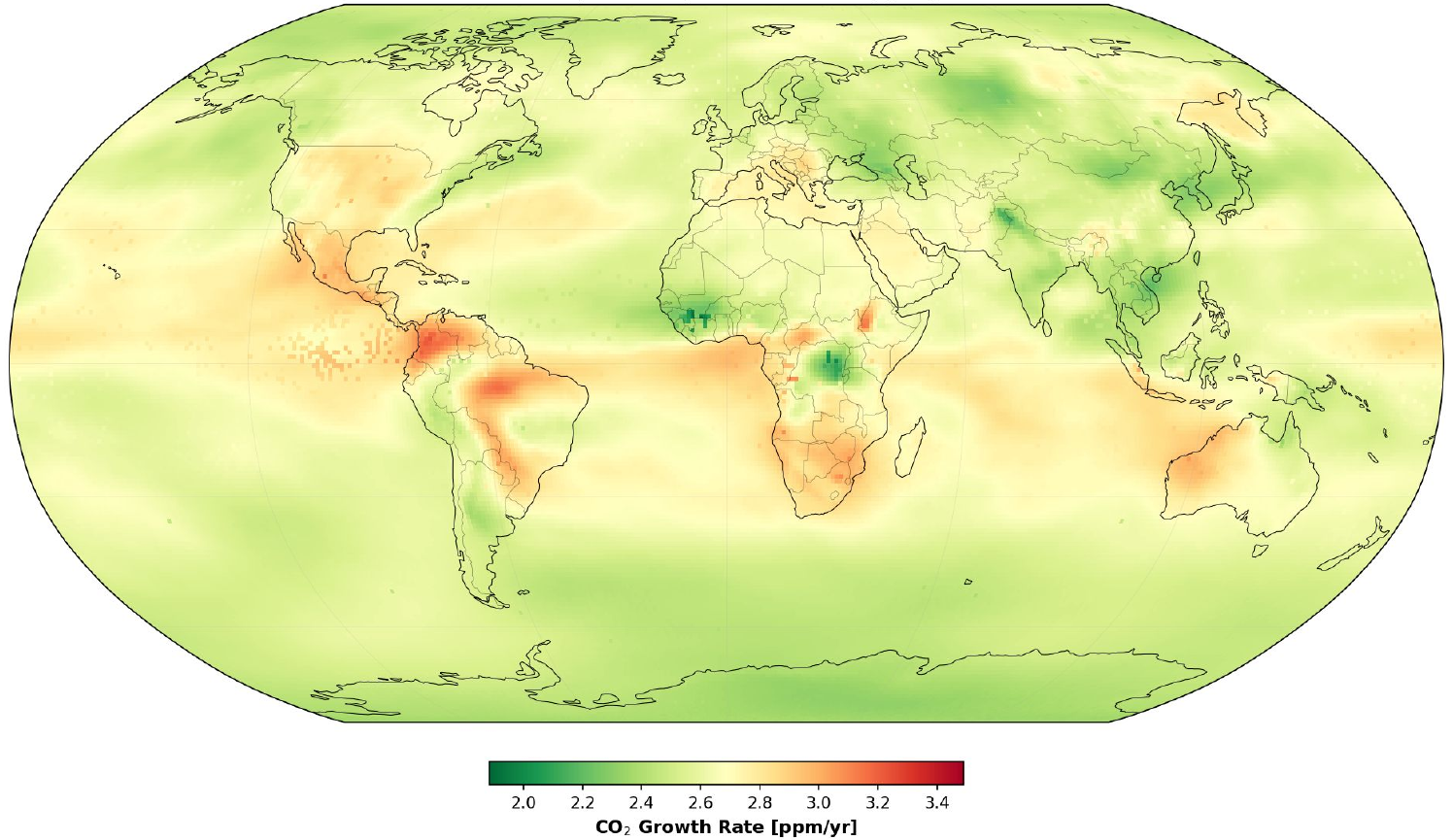}
    \caption{Global CO$_2$ growth rate in 2019 estimated from deseasonalized DLM-fitted \(\ce{XCO_2}\). Colors show annual growth rate in ppm\,yr\(^{-1}\); the global mean was 2.58 ppm\,yr\(^{-1}\). }
    \label{fig:data}
\end{figure}

\begin{itemize}

    \item DSg, annual CO$_2$ growth rates. We used the data from the CAMS greenhouse gas reanalysis which provides the column-averaged dry-air mole fraction of CO$_2$, denoted \(\ce{XCO_2}\), on a $1^\circ \times 1^\circ$ global grid. \(\ce{XCO_2}\) is available at 3-hourly temporal resolution from 2014 onward. On every grid point, we first calculated weekly averages, which were then fitted using a Dynamic Linear Model (DLM) implemented with the ``dlmhelper'' package developed in \cite{Hachmeister2024}. The DLM fit provided estimates of the \ce{XCO_2} levels at the same temporal resolution. Annual growth rates were calculated as the difference between deseasonalized annual means of the DLM fit for consecutive years, following the procedure described in \cite{MostafaviPak2025AnnualTCCON}.  In Fig.~\ref{fig:data} we show, as an example, the gridded \ce{CO_2} growth rate map for 2019.

    \item DSe, annual CO$_2$ emissions. The EDGARD dataset \citep{Crippa2025GHGEmissionsAllWorldCountries} provides the annual anthropogenic CO$_2$ emission, in a grid of $0.1^{\circ}\times0.1^{\circ}$, for the period  1970--2024, covering the whole globe. For CO$_2$, the dataset provides separate fossil CO$_2$ emissions, also referred to as IEA--EDGAR CO$_2$, and biofuel-combustion CO$_2$ emissions, also referred to as EDGAR \ce{CO2bio}. The fossil component includes fossil-fuel combustion and industrial/process sources, while the bio component refers to emissions from biofuel combustion. Large-scale biomass burning, forest fires, and land-use, land-use change and forestry sources and sinks are excluded from these \ce{CO_2} components. In this study, we summed the emissions onto a $1^{\circ}\times1^{\circ}$ grid cell and year to obtain the EDGAR CO$_2$ emission variable used in the analysis. 

    \item DSb, solar-induced chlorophyll fluorescence as a proxy for vegetation photosynthetic activity. We used the Global OCO-2 Solar-Induced Chlorophyll Fluorescence (GOSIF) dataset \citep{Li2019AData}, which provides global gridded solar-induced chlorophyll fluorescence (SIF) estimates. SIF is closely related to photosynthetic activity and is widely used as a proxy for gross primary productivity (GPP), but it is not a direct measure of net vegetation CO$_2$ uptake. GOSIF combines Orbiting Carbon Observatory-2 (OCO-2) SIF soundings with Moderate Resolution Imaging Spectroradiometer (MODIS) vegetation data and Modern-Era Retrospective Analysis for Research and Applications, Version 2 (MERRA-2) meteorological data. The product is available at $0.05^\circ \times 0.05^\circ$ spatial resolution and 8-day temporal resolution. The original product covered 2000--2017, and we used the currently available extended version covering 2000--2024. The 8-day SIF estimates were aggregated to annual values and onto a $1^{\circ}\times1^{\circ}$ grid for the present analysis.

    \item DSs, 7 months shifted southern oscillation index (SOI). The SOI is a standardized index based on the monthly normalized sea level pressure difference between Tahiti and Darwin (Australia), which is commonly used as an indicator of El Ni\~no--Southern Oscillation (ENSO) state \citep{Ropelewski1987SOI}. Although SOI is an atmospheric pressure index, it is closely related to ENSO-driven variability in tropical Pacific sea surface temperature and atmospheric circulation. Following \citep{buchwitz2018computation} who used a 7 months shifted SOI to analyse variations in the atmospheric CO$_2$ growth rate, we applied the same 7 months lag to the monthly SOI time series before using it in our analysis.

\end{itemize}

In order to use these three datasets in a combined analysis, we homogenized the datasets to a spatial resolution of $1^{\circ}\times1^{\circ}$, a temporal resolution of $1$ year, and covering the period 2015--2024. These coarse-grainings were obtained by spatial averaging over the $1^{\circ}\times1^{\circ}$ grid, and the temporal average over the calendar year.

\begin{table}[h]
\caption{Summary of datasets used in the analysis.}
\centering
\begin{threeparttable}
\footnotesize

\setlength{\tabcolsep}{4pt}
\renewcommand{\arraystretch}{1.15}

\begin{tabular}{@{}
>{\RaggedRight\arraybackslash}p{1.6cm}
>{\RaggedRight\arraybackslash}p{3.5cm}
>{\RaggedRight\arraybackslash}p{2.7cm}
>{\RaggedRight\arraybackslash}p{2.4cm}
>{\RaggedRight\arraybackslash}p{4.1cm}
@{}}

\toprule
\textbf{Source} &
\textbf{Variable used} &
\makecell[l]{\textbf{Spatial}\\\textbf{resolution}} &
\makecell[l]{\textbf{Temporal}\\\textbf{resolution}} &
\makecell[l]{\textbf{Processing for}\\\textbf{this study}} \\
\midrule

CAMS\tnote{a} &
XCO$_2$-derived annual CO$_2$ growth rate &
$ 1^\circ \times 1^\circ$ global grid &
3-hourly, from 2003 &
Weekly averaging, DLM fitting, and annual growth-rate calculation. \\[0.35em]

EDGAR\tnote{b} &
Fossil and biofuel-combustion CO$_2$ emissions &
$0.1^\circ \times 0.1^\circ$ global grid &
Annual, 1970--2024 &
Fossil and biofuel-combustion components summed by grid cell and year. \\[0.35em]

GOSIF\tnote{c} &
Solar-induced chlorophyll fluorescence (SIF) &
$0.05^\circ \times 0.05^\circ$ global terrestrial grid &
8-day, 2000--2024 &
Aggregated to annual values. \\[0.35em]

SOI\tnote{d} &
Southern Oscillation Index shifted by 7 months &
Global climate index &
Monthly &
Applied 7-month lag. \\

\bottomrule
\end{tabular}

\begin{tablenotes}[flushleft]
\footnotesize
\item[a] CAMS processing and annual growth-rate calculation follow \citet{Hachmeister2024} and \citet{MostafaviPak2025AnnualTCCON}.
\item[b] EDGAR refers to the EDGAR\_2025\_GHG dataset; data and gridding methodology are described in \citet{Crippa2025GHGEmissionsAllWorldCountries} and \citet{Crippa2024InsightsV8.0}.
\item[c] GOSIF is described in \citet{Li2019AData}.
\item[d] SOI definition and lagged use follow \citet{Ropelewski1987SOI} and \citet{buchwitz2018computation}.
\end{tablenotes}

\label{tab:datasets}
\end{threeparttable}
\end{table}

\section{Methods}\label{sec:Methods}

\begin{figure}
    \centering
    \includegraphics[width=0.7\linewidth]{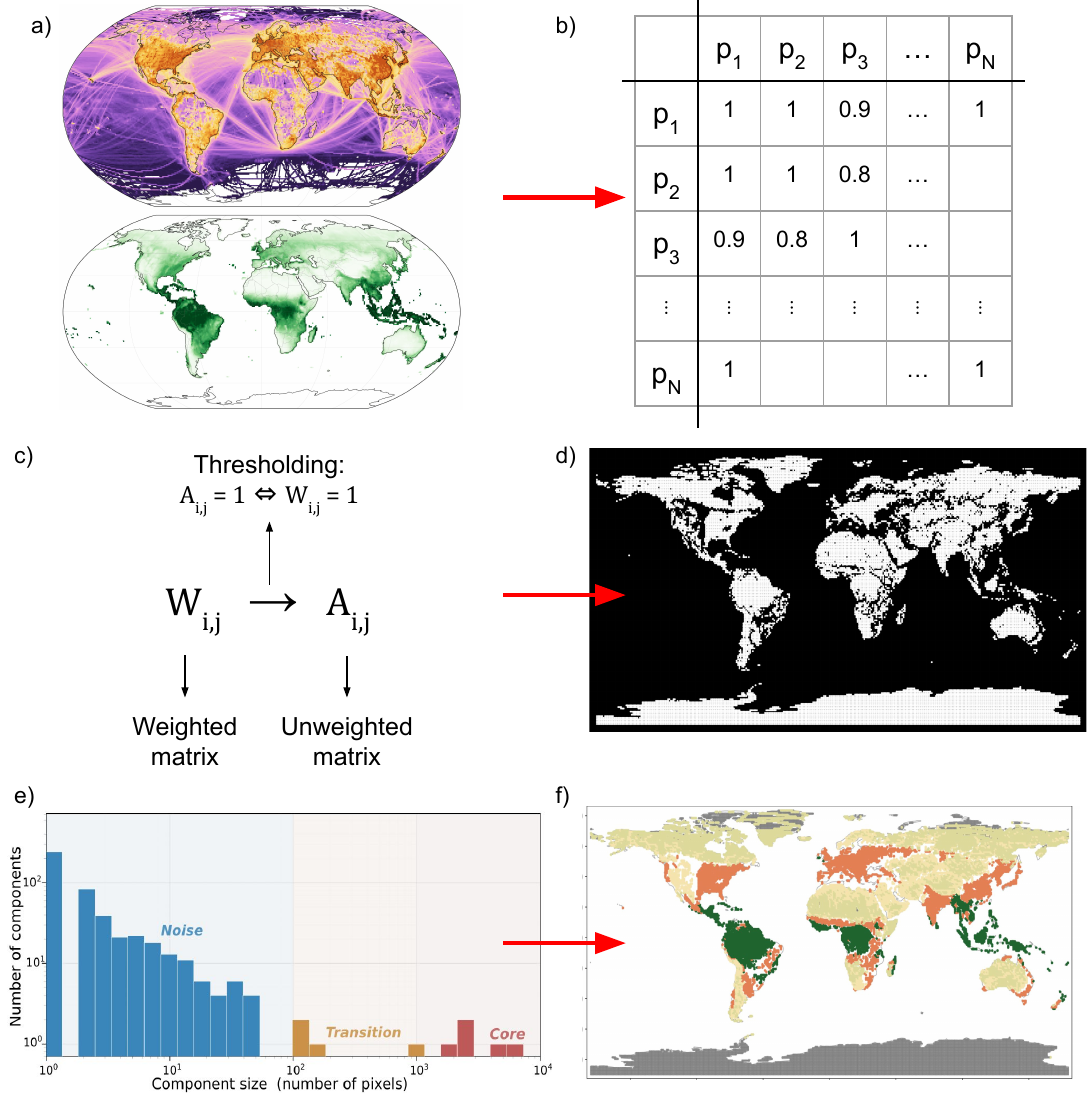}
    \caption{Summary of the persistence analysis of the clusters. a) Snapshots of the input data , DSe and DSb, from 2019. b) Illustration of persistence matrix, $W_{i,j}$, representing the frequency that pixels $p_i$ and $p_j$ belonged to the same cluster, obtained from k-clusters algorithm applied to the input data. c) Thresholding process to obtain $A_{i,j}$, where $A_{i,j}=1\iff W_{i,j}=1$.  d) Pixels removed (in black) by the thresholding process.
    e) Illustration of the size distribution of the connected components of the matrix $A_{i,j}$. It should be noted that the sizes of the persistent clusters, expressed in grid cells, do not translate directly to an area on Earth's surface. For that, a latitude-dependent correction should be applied to each cell. f) Clusters obtained from this analysis when keeping 5 largest components of $A_{i,j}$.. }
    \label{fig:methods}
\end{figure}
\subsection{Linear Regression}\label{sec:methods:reg}

We fitted independent pixel-wise linear regression models at each land grid cell
$r=(\mathrm{lat},\mathrm{lon})$, using annual observations from 2015--2024 as
temporal samples. As an initial diagnostic specification, we examined whether
local annual atmospheric CO$_2$ growth rates were statistically associated
with local anthropogenic emissions and biospheric activity:

\begin{equation}\label{eq:LR}
g(t,r)
=
\beta_0(r)
+
\beta_E(r)E_z(t,r)
+
\beta_B(r)B_z(t,r)
+
\varepsilon(t,r),
\end{equation}

where $g(t,r)$ denotes the annual atmospheric CO$_2$ growth rate, $E(t,r)$ denotes anthropogenic emissions (DSe), and $B(t,r)$ denotes biospheric activity represented by GOSIF (DSb). The analysis was restricted to land grid cells because GOSIF provides a vegetation-related signal and is not defined over ocean regions. The predictors were standardized prior to fitting—by subtracting the temporal mean and dividing by the temporal standard deviation—so that the estimated coefficients could be compared on a common scale. In this formulation, $\beta_E(r)$ represents the local emission-related association with atmospheric
CO$_2$ growth, while $\beta_B(r)$ represents the local biosphere-related association. The coefficients are therefore interpreted as statistical associations rather than direct source--sink attribution estimates. The intercept $\beta_0(r)$ represents the fitted baseline component of the local response, while unexplained processes are contained in the residual term
$\varepsilon(t,r)$.

As sensitivity analyses, we also tested a background-corrected response, intended to remove the globally coherent, interannual ENSO-driven signal from the target and so better isolate the local response to the surface drivers,

\begin{equation}
g^{\prime}(t,r) = g(t,r) - g_{\mathrm{SPO}}(t),
\end{equation}

where $g_{\mathrm{SPO}}(t)$ is the South Pole annual growth rate, and models using year-to-year changes in the surface drivers:

\begin{equation}
g^{\prime}(t,r)
=
\beta_0(r)
+
\beta_E(r)\Delta E_z(t,r)
+
\beta_B(r)\Delta B_z(t,r)
+
\varepsilon(t,r),
\qquad t=2016,\ldots,2024.
\end{equation}

Here, $\Delta E_z(t,r)=E_z(t,r)-E_z(t-1,r)$ and
$\Delta B_z(t,r)=B_z(t,r)-B_z(t-1,r)$. These emissions--GOSIF models were retained as diagnostic analyses and are reported in the Supplementary Information.

Because annual atmospheric CO$_2$ growth also contains large-scale interannual climate variability, we fitted a second pixel-wise regression specification including the Southern Oscillation Index (SOI) as an ENSO-related predictor. Following previous study by \citep{buchwitz2018computation},the SOI time series was shifted by a 7-month  lag and then averaged to annual values. Unlike emissions and GOSIF, SOI is not spatially gridded; it varies only in time and is therefore used as a common annual predictor across all land grid cells. However, its regression coefficient is estimated separately at each grid cell, allowing the association between atmospheric CO$_2$ growth and SOI variability to vary spatially.
 For each land grid cell, we fitted

\begin{equation}\label{eq:SOI_emission}
g(t,r)
=
\beta_0(r)
+
\beta_{\mathrm{SOI}}(r)\,\mathrm{SOI}_z(t)
+
\beta_E(r)E_z(t,r)
+
\varepsilon(t,r).
\end{equation}

Here, $\mathrm{SOI}_z(t)$ is the standardized lagged annual SOI index and $E_z(t,r)$ is the standardized anthropogenic-emission field.  The coefficient $\beta_{\mathrm{SOI}}(r)$ represents the local association between annual atmospheric CO$_2$ growth and ENSO-related interannual variability, while $\beta_E(r)$ represents the local emission-related association. For this model, we used the raw atmospheric CO$_2$ growth rate
rather than the South Pole-corrected response, because SOI was included explicitly to represent part of the large-scale interannual variability. All models were fitted using both ordinary least squares and ridge regression; the two methods yielded consistent results throughout, so we report ordinary least and report the ridge results in the Supplementary. For ridge regression, the penalization parameter $\lambda$ was selected  independently at each grid cell by leave-one-year-out cross-validation, choosing the value that minimized the cross-validated prediction error. Predictors were standardized before fitting so that coefficient magnitudes were comparable and ridge regularization was not affected by differences in predictor scale.

\subsection{Clustering algorithm: K-means}\label{sec:methods:kmeans}

We employ the simple and widely used K-means algorithm to partition data points into $k$ groups \cite{Hartigan1979KMeans}. K-means was chosen because our dataset is numerical and our objective is to identify natural groups in the data without using predefined labels, which corresponds to an unsupervised learning task. The method is also well suited for large datasets with many observations, as it is computationally efficient and easy to interpret. For a fixed number of clusters $k$, K-means assigns $N$ observations $x_i \in \mathbb{R}^d$, with $i = 1, \ldots, N$, to $k$ disjoint clusters $\{C_1, C_2, \ldots, C_k\}$ by minimizing the total within-cluster distance. More precisely, it finds cluster centers $\mu_j$ that minimize $\sum_{j=1}^{k} \sum_{x \in C_j} \|x - \mu_j\|^2$, where $\|\cdot\|$ denotes the Euclidean norm. We then use several complementary methods to determine the optimal number of clusters $k^*$, as detailed in Sec.~3 of the Supplementary Information.

We then translate $W$ to a binary matrix $A $ by setting $A_{ij} = 1$ if $W_{ij} =1$ and $0$ otherwise, so that the matrix $A$ links the groups of grid cells that remain clustered together consistently across the ten-year period, regardless of the number of clusters.  This strict rule grants $A$ a block structure, effectively performing a second clustering step that separates the core of the original clusters (the pixels that are always clustered together) from those with a more uncertain classification. As we will show in the results, these latter groupings can also be informative. This process is schematized in Fig.~\ref{fig:methods}. 

\subsection{Linear Regression on aggregated data}

We use the clusters from Section \ref{sec:methods:kmeans} to aggregate data for a second round of linear regression models. For each  persistent cluster $c$ and year $t$, we compute the unweighted mean growth rate, mean emission, and mean GOSIF index over all the $1^o$ grid cells assigned to cluster $c$ in year $t$: 
$$\bar{X}_c(t) = \frac{1}{|S_c(t)|} \sum_{(i,j) \in S_c(t)} X(i, j, t)$$ where $S_c(t)$ denotes the set of grid cells $(i,j)$ belonging to component $c$ in year $t$, and $X$ is a vector of the variables being averaged.
Grid cells that did not appear in the persistent clusters were excluded from the analysis.
This yielded a $10$-year time series of cluster-mean values for each variable and component.

We then fit two ordinary least squares (OLS) regression models using the cluster mean time series as observations, similar to the method from Section \ref{sec:methods:reg}:

$$y_c(t) = \beta_0 + \beta_E , E_z^c(t) + \beta_2 , P_z(t) + \varepsilon(t)$$

where $y_c(t)$ is the component-mean raw CO$_2$ growth rate, $E_z^c(t)$ is the standardized cluster-mean emission, and $P_z(t)$ is the second predictor, which differs between the two models:
\begin{itemize}
    \item \textbf{Model SOI}: $P_z = \text{SOI}_z(t)$, the global annual SOI index lagged 7 months and standardized over 2015–2024.
    \item \textbf{Model GOSIF}: $P_z = \text{GOSIF}_z^c(t)$, the standardized cluster-mean GOSIF.
\end{itemize}

Both predictors were standardized independently for each component by subtracting the temporal mean and dividing by the temporal standard deviation over the 10-year period.

The models' performance was assessed using two complementary metrics. The in-sample coefficient of determination ($R^2$) which measures the proportion of temporal variance in the cluster-mean growth rate explained by the fitted models; and the leave-one-year-out (LOO) cross-validated $R^2$ (LOO$R^2$). The LOO$R^2$ was used to estimate the predictive skill of the model which was necessary because of the small number of observations available for training. The LOO$R^2$ was computed analytically using the hat-matrix (also, known as projection matrix) shortcut:

$\hat{y}_{-i} = y_i - \frac{y_i - \hat{y}_i}{1 - h_i}$

where $\hat{y}_{-i}$ represents the predicted value of the omitted observation,  and $h_i$ is the leverage of observation $i$ corresponding to the $i$-th diagonal element of the projection matrix. This avoids the computational cost of refitting the model ten times while yielding exact LOO predictions. 

\section{Results}\label{sec:Results}

\subsection{Gridded prediction of CO$_2$ growth rate}
\begin{figure}[h!]
    \centering
    \includegraphics[width=0.99\linewidth]{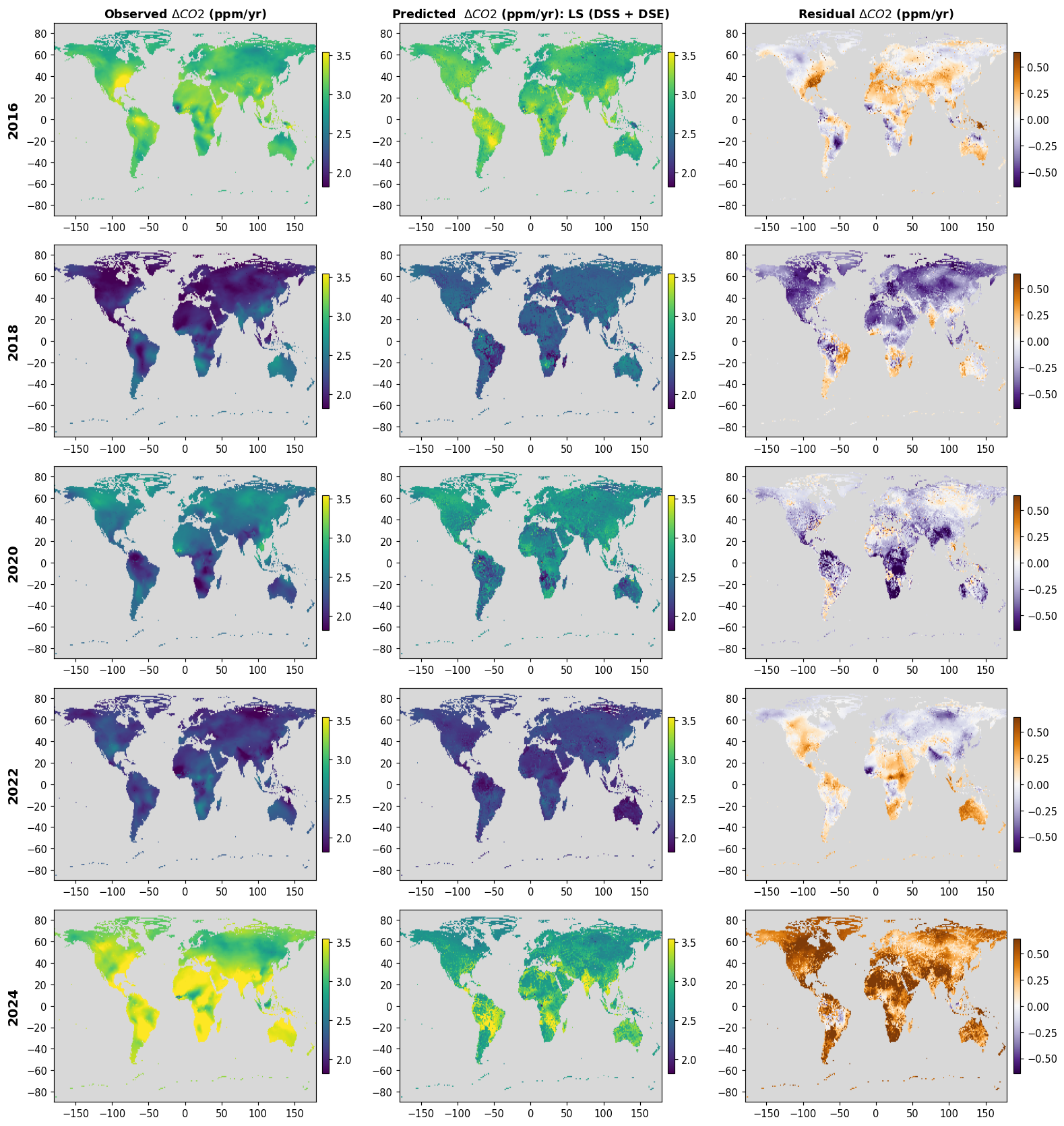}
    \caption{Observed and predicted CO$_2$ growth rates for selected years. \textit{First column:} observed growth rates from the [DSg] dataset. \textit{Second column:} predicted growth rates from an ordinary least squares (OLS) linear regression on local emission [DSe] and the lagged, global annual SOI index [DSs], which agrees closely with the observed rates despite the model's simplicity. \textit{Third column:} residuals (observed $-$ predicted). 2016 and 2024 are shown as examples of strong El Ni\~no years, while 2020 is shown as an example of an anomalous year. Supplementary Fig.~1 repeats this regression using ridge regression with an optimal per-cell penalization parameter, which yields similar results, and also shows the regression using the year-to-year change in emission, $\Delta E(r,t)$, in place of its absolute value. Supplementary Fig.~2 shows the corresponding driver coefficient maps $\beta_i(r)$ together with the goodness of fit $R^2(r)$.}
    \label{fig:obs_vs_pred_levels}
\end{figure}

First, we present the results obtained from the linear regression analysis (Eq.~\ref{eq:LR}), reporting the least squares method. Fig.~\ref{fig:obs_vs_pred_levels} shows the observed levels of CO$_2$ growth rates  (DSg) in the land regions of the globe, their predicted values based on the standardized values from the DSe and DSs datasets, and their residuals showing the observed - predicted values at every pixel. We show these results every 2 years in the period 2016--2024. This figure highlights how heterogeneous, in space and time, the CO$_2$ growth rates are. Here we can also observe the impact that ENSO has on observed rates, as during 2016 and 2024 strong El Niño events occurred, coinciding with generalized high positive growth rates throughout the globe, while during 2018 and 2022 correspond to La Niña years and negative rates are observed \citep{Lan2026NOAACo2Trends}. For 2020, which was a neutral year for ENSO, the spatial distribution of residuals is primarily positive for the northern hemisphere, while primarily negative rates are observed for the southern hemisphere. 

During these ENSO events the CO$_2$ growth rates are expected to  increase (during El Niño) and decrease (during La Niña) than normal. From Fig. \ref{fig:obs_vs_pred_levels} the prediction of our model tends to have the lowest residuals in the tropical forest during the El Niño years . However in many other regions of the world the model could not capture the increase in the growth rate in regions such as East North America, North Africa and Southern Europe in 2016. The underestimation is even more pronounced in 2024 where almost all regions of the world we observe it other than the tropical regions. The predictions for 2020, are closer to zero in many regions, which is considered a neutral ENSO year. However, we expect lower growth rates in highly populated areas due to reduction in fossil fuel emissions.  

Before settling on this model, we also tested local predictors based on gridded biospheric activity. A regression of the growth-rate anomaly on local anthropogenic emission and local GOSIF (a proxy for biospheric photosynthetic activity) yielded a markedly worse fit, with higher residuals and lower $R^2(r)$ almost everywhere (Supplementary Section 2). This poor performance suggests that grid-cell CO$_2$ growth-rate anomalies are not controlled by local emissions and local biospheric activity alone; a natural next step would be to replace these purely local predictors with spatially convolved predictors that account for atmospheric transport and regional mixing. The regression reported in Fig.~\ref{fig:obs_vs_pred_levels}, performed substantially better overall, with lower residuals and higher $R^2(r)$, although the improvement remains strongly location-dependent.  In the supplementary material we report the regression using local GOSIF and emission in more detail (Supplementary section 2).

\begin{figure}[h!]
    \centering
    \includegraphics[width=0.99\linewidth]{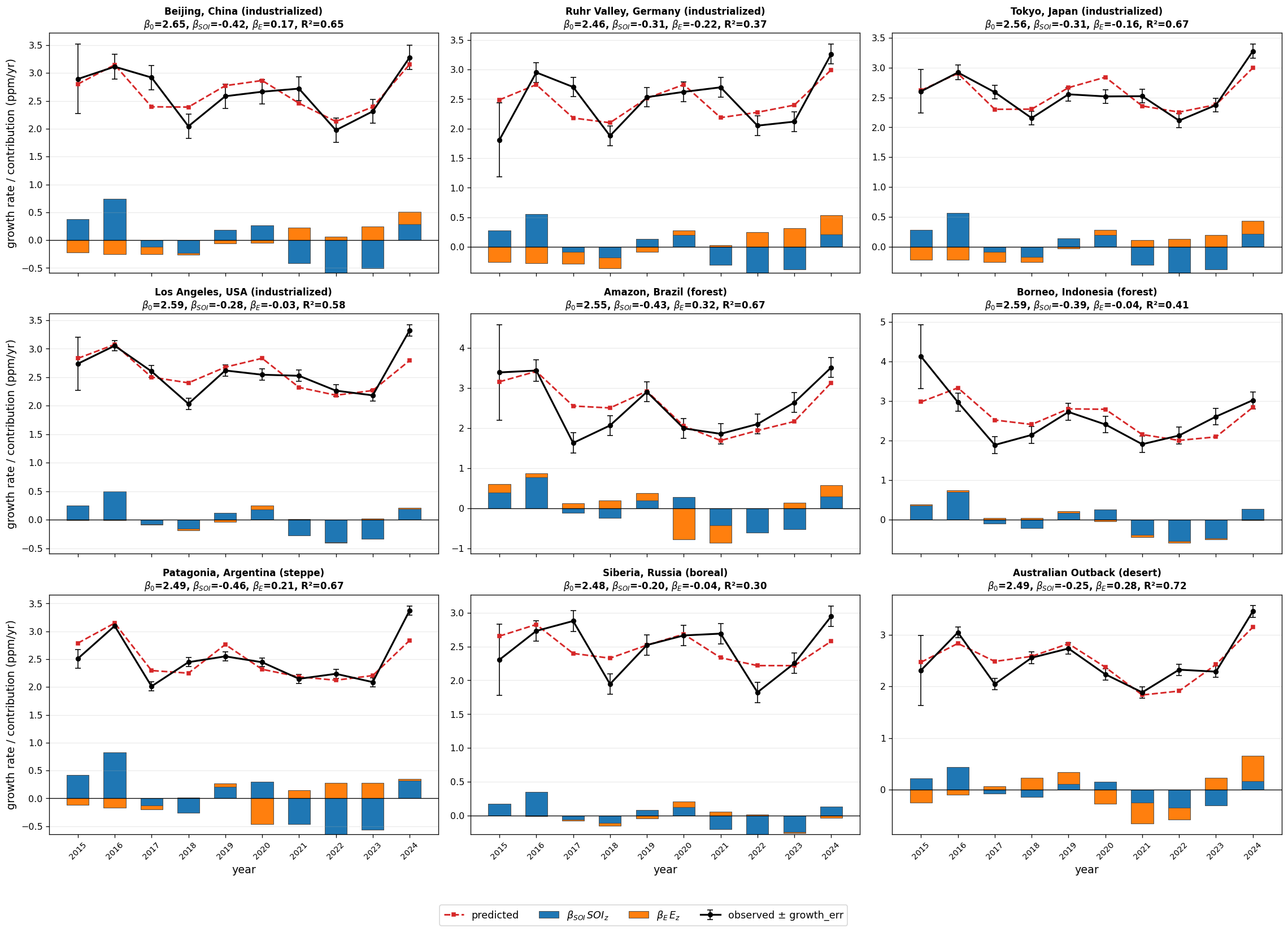}
    \caption{Year-by-year evolution of the observed (with error bars) and predicted CO$_2$ growth rates from the local linear regression at 9 locations spanning different land-cover categories (industrialized cities, forest, desert, boreal, and steppe; see Fig.~\ref{fig:comp_map_t1} for their geographical positions). Each panel reports the local regression coefficients $\beta_i(r)$ and the goodness of fit $R^2(r)$. The lower part of each panel shows the local contribution of each driver, $\beta_S(r)\cdot\mathrm{SOI}_z(t)$ and $\beta_E(r)\cdot E_z(t,r)$; the remaining, undrawn contribution is the baseline term $\beta_0(r)$, which accounts for most of the absolute growth rate. At most locations and time the predicted growth rate falls within the error bars of the observed values.}
    \label{fig:ts_locations}
\end{figure}

In Fig.~\ref{fig:ts_locations}, we show the evolution in the different years for 8 different grid points that include large industrial sources and cities (Beijing, Ruhr Valley, Tokyo and Los Angeles). A pixel from the amazon forest and Borneo island representing the equatorial rain forest. Patagonia desert, Siberia and Australian desert to show variations in growth rate in different low emission regions. In general, the agreement between model predictions and observed growth rates is fairly good, with specific spatial locations where the model fails to predict the actual growth rate. For example, the prediction in Siberia, which presents the worst agreement between observation and prediction, fails due to the lack of direct driver influence, as in this region there is almost no human or vegetation activity, and the evolution of the CO$_2$ growth is governed by its transport from other regions. There are also moments in time where the prediction also fails, mainly due to events that greatly affected the drivers. For example, biogenically driven region (like the Amazon and Borneo) fail by a large amount in the 2017 prediction,  because it is severely impacted by the strong El Ni{\~n}o event of 2016, which had a rebound effect the following year. It is this rebound effect that our model fails to capture. 
Beside these specific moments, the best agreement is observed in locations that have a strong influence of the drivers, like cities (Beijing, Tokyo and Los Angeles) and tropical forest (Amazon and Borneo). In this figure we also show the contributions of each driver to the prediction of CO$_2$ growth (colored bars), which can allow us to uncover issues in our model.
For example, during the COVID-19 pandemic (2020), the human emissions in cities was significantly reduced \citep{Nicolini2022UrbanCO2Lockdown}. However, our model fails to display this effect, as in Ruhr Valley, Tokyo, and Los Angeles, the contribution from emission is increased respect to the previous year, while Beijing maintains similar levels. In this case, the influence of ENSO could still be strong in the cities, masking the actual effect of human emissions.

\subsection{Clustering regions based on the \texorpdfstring{\ce{CO2}}{CO2} drivers}

\begin{figure}[]
    \centering
    \includegraphics[width=0.99\linewidth]{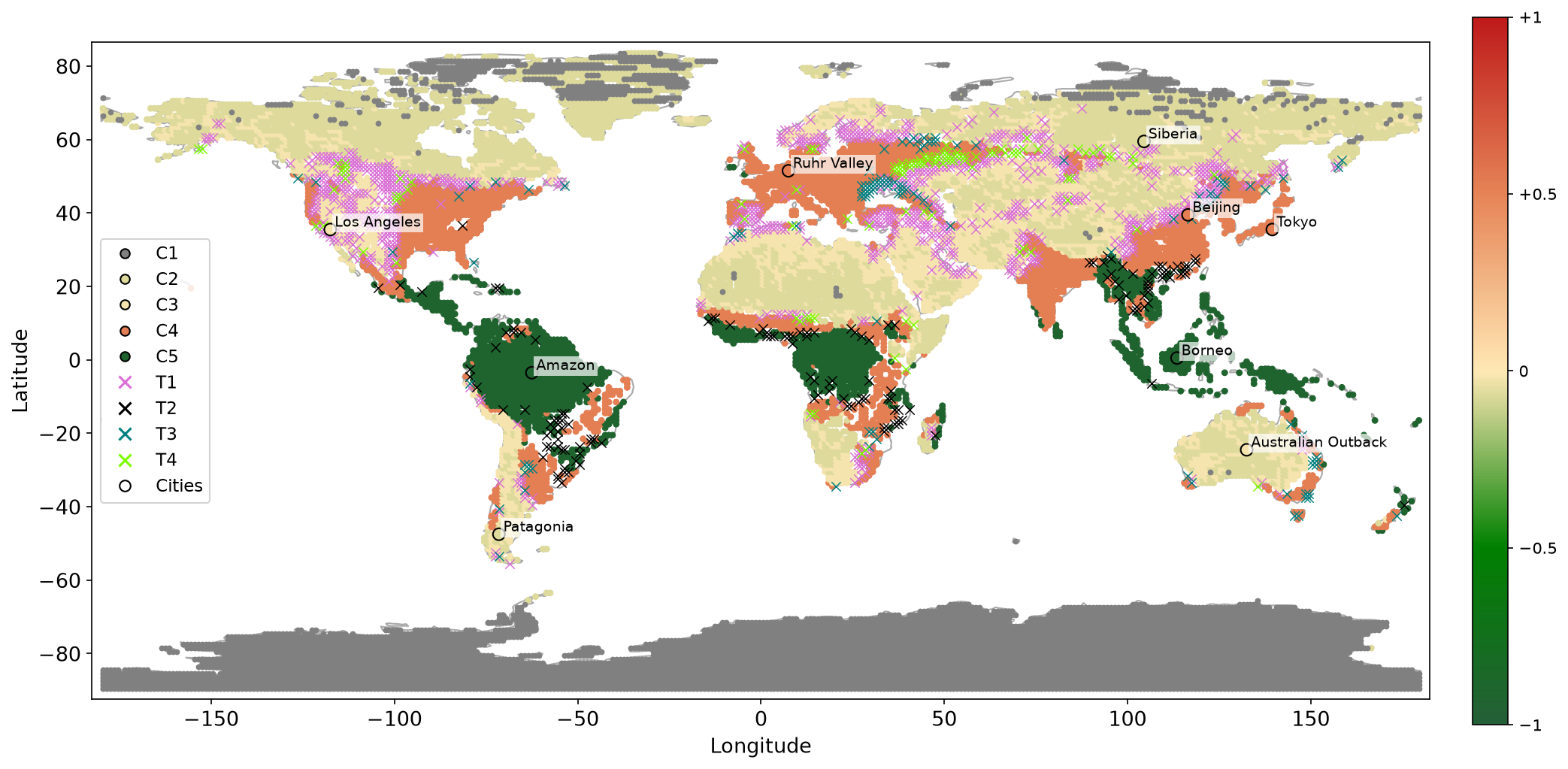}
    \caption{Five largest persistent clusters over the ten-year period (C1--C5). They are plotted using a normalized color scaling that indicates if each cluster is more anthropogenically driven (positive values), or biogenically driven (negatives values), while values close to zero correspond to low influence from the drivers. Here we also show 4 transitional clusters (T1--T4) as colored crosses. Unlike isolated classification noise, these transitional clusters exhibit coherent temporal behavior, meaning the entire group oscillates between two primary clusters over time. T1 corresponds to transitions between C3 and C4, while T2--T4 to transition between C4 and C2 and C3.  Gray corresponds to no driver presence. }
    \label{fig:comp_map_t1}
\end{figure}

\begin{figure}[]
    \centering
    \raisebox{-0.5\height}{\includegraphics[width=.8\linewidth]{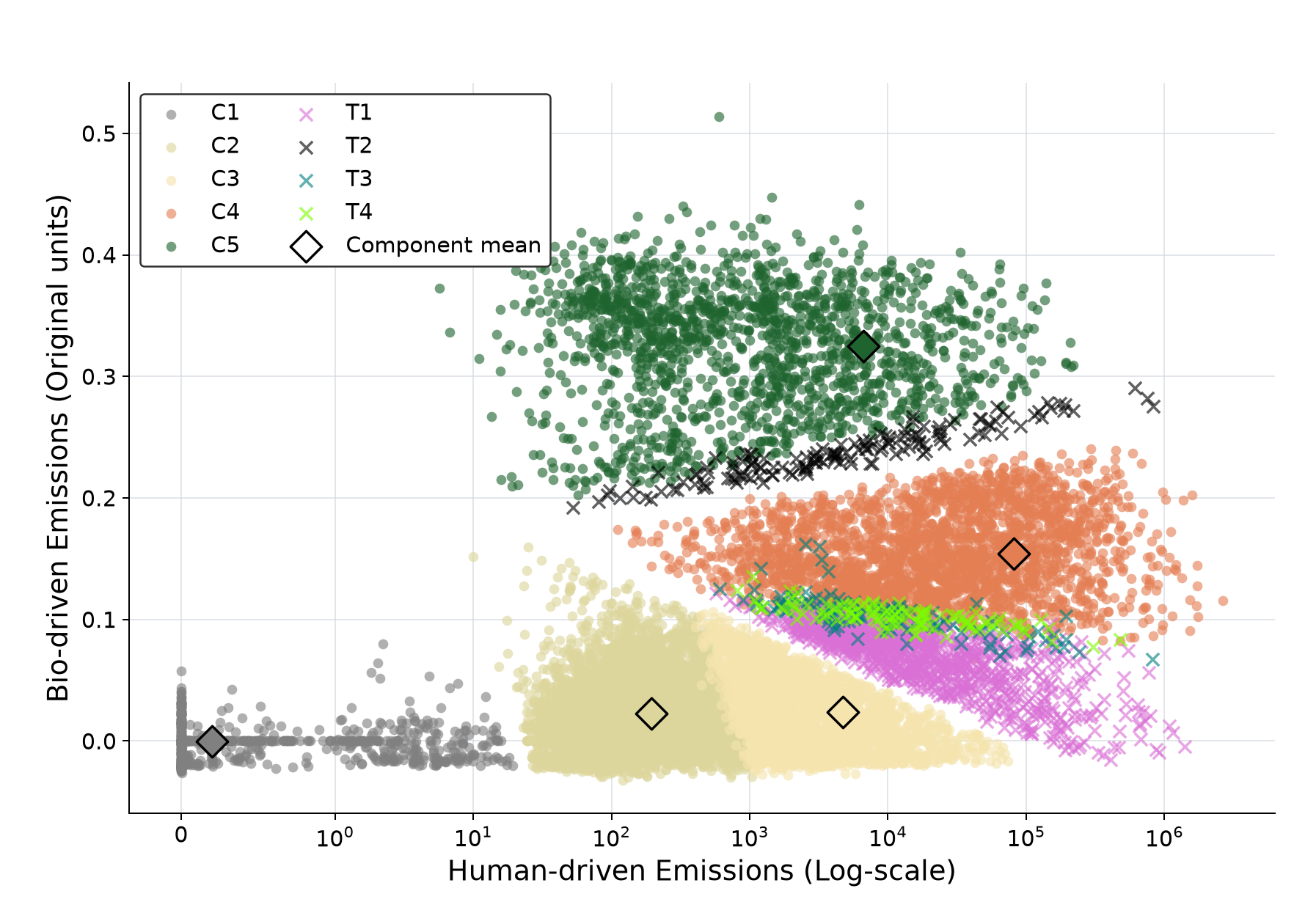}}
    \caption{Mean value of human-driven emissions (DSe) and bio-driven emissions (DSb) in the different clusters obtained from the persistence analysis and shown in Fig.~\ref{fig:comp_map_t1}. These values are obtained by first averaging each pixel value in the 10 years, and then averaging in each cluster's persistent pixels. } 
    \label{fig:cluster_means}
\end{figure}

In this section, we present the classification of different regions of the globe based on their values in the DSe and DSb datasets, using the main components identified from the persistence analysis. The interpretation combines three complementary elements: the geographical organization of the persistent components in Fig.~\ref{fig:comp_map_t1}, their distribution in the (DSe, DSb) variable space in Fig.~\ref{fig:cluster_means}, and the physical interpretation summarized in Table~\ref{tab:clusters_descriptions}. The map indicates where each component is located on the globe, while the (DSe, DSb) projection represents each pixel by its 10-year mean values. In this projection, DSb is displayed on a logarithmic scale to better resolve the separation between components in the low-DSb range. This joint reading allows each cluster to be interpreted not only from its geographical location, but also from its relative anthropogenic and biogenic signatures.

Cluster C1 represents the lowest-activity end of the classification, corresponding to regions where carbon-cycle activity is minimal. Shown in gray in Fig.~\ref{fig:comp_map_t1}, this cluster is mainly located in polar cryospheric regions, such as Antarctica and Greenland, where direct anthropogenic activity is very limited and biospheric activity remains weak. This low-activity regime is also evident in the (DSe, DSb) space shown in Fig.~\ref{fig:cluster_means}, where C1 occupies the lowest range of both variables, supporting its characterization in Table~\ref{tab:clusters_descriptions} as carbon inactive.

Moving away from this nearly inactive regime, clusters C2 and C3 correspond to regions where the biogenic signal remains weak, while the anthropogenic contribution separates the two clusters. C2, shown in pale khaki in Fig.~\ref{fig:comp_map_t1}, includes regions such as Alaska, Siberia, and the Sahara. In the (DSe, DSb) space, it remains close to C1, with low DSb values and very low DSe values, indicating limited natural productivity together with weak direct anthropogenic influence. This supports its interpretation as a carbon-limited regime associated with arid and boreal regions. C3, shown in pale yellow, occupies a comparable low-DSb range, but with higher DSe values than C2. This shift indicates a stronger anthropogenic contribution while the biogenic signal remains limited, which is consistent with regions such as the central United States and motivates its interpretation as an intermediate carbon-transition regime associated with anthropogenic drylands.

The classification then reaches two high-activity regimes, where both DSe and DSb are relatively large but where the dominant contribution differs. C4, shown in orange in Fig.~\ref{fig:comp_map_t1}, is characterized by the strongest anthropogenic signature, with higher DSe values than C5. It is therefore associated with densely populated and highly urbanized regions, and is interpreted in Table~\ref{tab:clusters_descriptions} as an anthropogenic core with carbon-source behavior. In contrast, C5, shown in dark green, is characterized by the strongest biogenic signature, with the highest DSb values in the (DSe, DSb) space. This cluster is mainly associated with tropical and highly productive biospheric regions, such as the Amazon, and is interpreted as an active biosphere with carbon-sink behavior. Together, C4 and C5 form the two main high-activity endpoints of the classification, separating regions dominated by anthropogenic emissions from regions dominated by biospheric activity.

The additional components T1--T4 refine this structure by highlighting intermediate persistent regions that are not included in the summary of Table~\ref{tab:clusters_descriptions}. Their presence in both Fig.~\ref{fig:comp_map_t1} and Fig.~\ref{fig:cluster_means} shows that the transitions between the five main clusters are not abrupt. In particular, the orchid (T1), teal (T3), and bright-green (T4) components occupy intermediate positions between C3 and C4 in the (DSe, DSb) space, indicating a gradual transition from weakly influenced regions toward regions with a stronger anthropogenic contribution. The black component (T2) is positioned between C4 and C5, separating the two high-activity regimes associated respectively with anthropogenic and biogenic dominance. These additional components therefore preserve the broader organization of the persistence-based classification and help connect the main Earth system regimes represented by C1--C5.

\begin{table}[t]
\centering
\begin{threeparttable}
\footnotesize

\setlength{\tabcolsep}{4pt}
\renewcommand{\arraystretch}{1.15}

\begin{tabular}{@{}
>{\centering\arraybackslash}p{1.0cm}
>{\RaggedRight\arraybackslash}p{1.8cm}
>{\RaggedRight\arraybackslash}p{3.0cm}
>{\RaggedRight\arraybackslash}p{3.0cm}
>{\RaggedRight\arraybackslash}p{3.1cm}
>{\RaggedRight\arraybackslash}p{3.0cm}
@{}}

\toprule
\textbf{Cluster} &
\textbf{Color} &
\textbf{Example region} &
\textbf{Driver} &
\textbf{Earth system theme} &
\textbf{Carbon cycle focus} \\
\midrule

C1 &
Gray &
Antarctica / Greenland &
None &
Cryosphere &
Carbon inactive \\[0.35em]

C2 &
Pale khaki &
Alaska / Siberia / Sahara &
Low biogenic influence, close no to anthropogenic influence &
Arid atmosphere / Boreal regions &
Carbon limited \\[0.35em]

C3 &
Pale yellow &
Central US &
Low influence from both &
Anthropogenic drylands &
Carbon transition \\[0.35em]

C4 &
Orange &
New York &
Anthropogenic &
Anthropogenic core &
Carbon source \\[0.35em]

C5 &
Dark green &
Amazon &
Biogenic &
Active biosphere &
Carbon sink \\

\bottomrule
\end{tabular}

\caption{Characteristics of the different clusters.}
\label{tab:clusters_descriptions}
\end{threeparttable}
\end{table}

\subsection{\texorpdfstring{\ce{CO2}}{CO2} growth rate trend analysis on the persistent clusters}

\begin{figure}
\centering
\begin{subfigure}[b]{.3\linewidth}
\includegraphics[width=\linewidth]{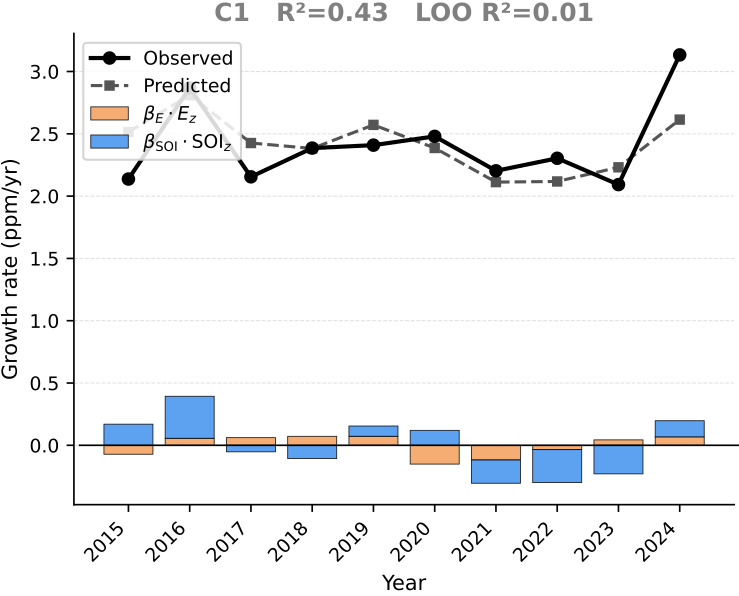}
\end{subfigure}
\begin{subfigure}[b]{.30\linewidth}
\includegraphics[width=\linewidth]{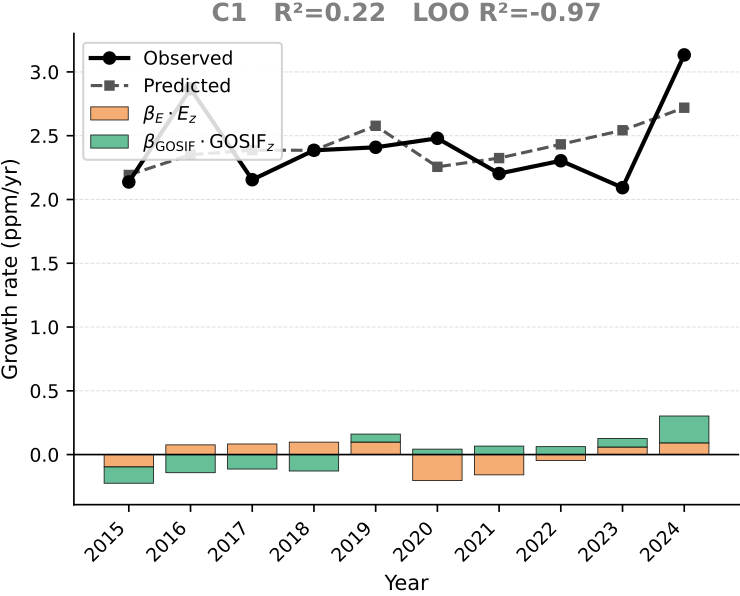}
\end{subfigure}

\begin{subfigure}[b]{.30\linewidth}
\includegraphics[width=\linewidth]{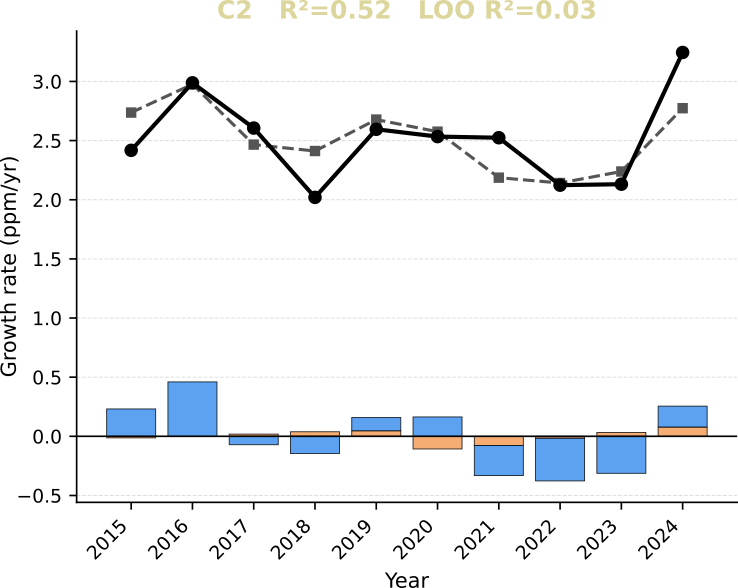}
\end{subfigure}
\begin{subfigure}[b]{.30\linewidth}
\includegraphics[width=\linewidth]{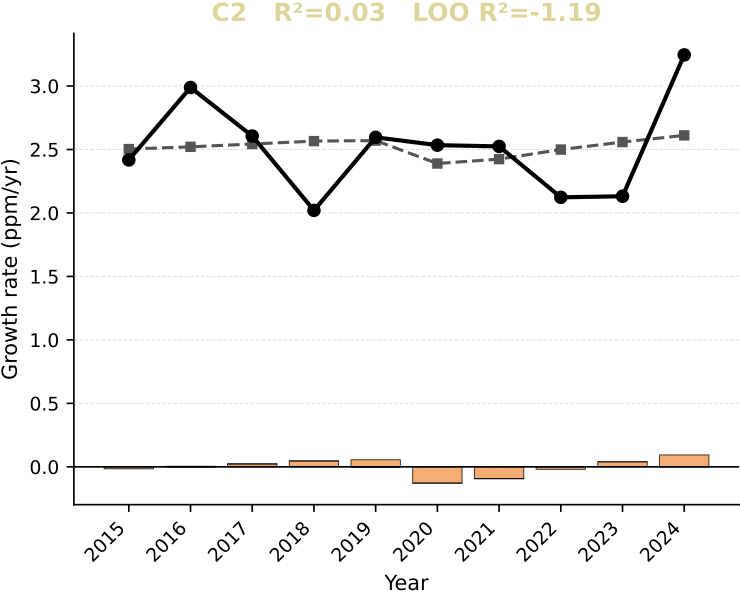}
\end{subfigure}

\begin{subfigure}[b]{.30\linewidth}
\includegraphics[width=\linewidth]{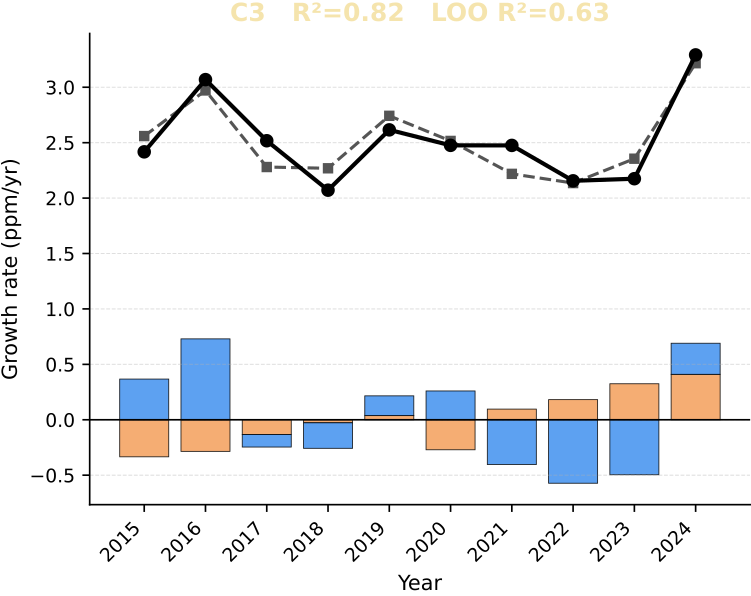}
\end{subfigure}
\begin{subfigure}[b]{.30\linewidth}
\includegraphics[width=\linewidth]{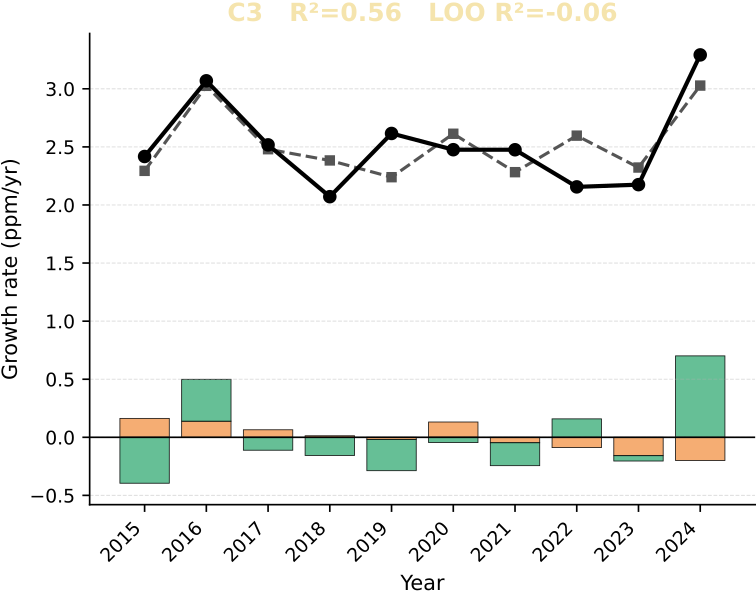}
\end{subfigure}

\begin{subfigure}[b]{.30\linewidth}
\includegraphics[width=\linewidth]{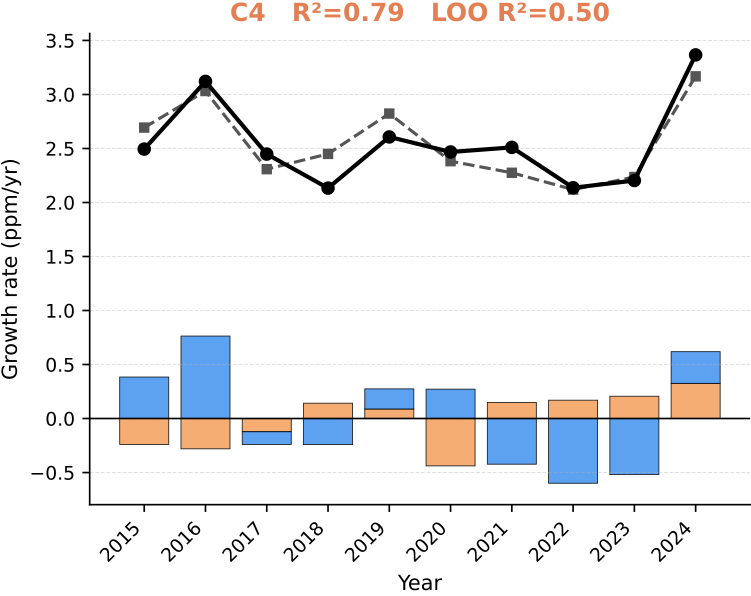}
\end{subfigure}
\begin{subfigure}[b]{.30\linewidth}
\includegraphics[width=\linewidth]{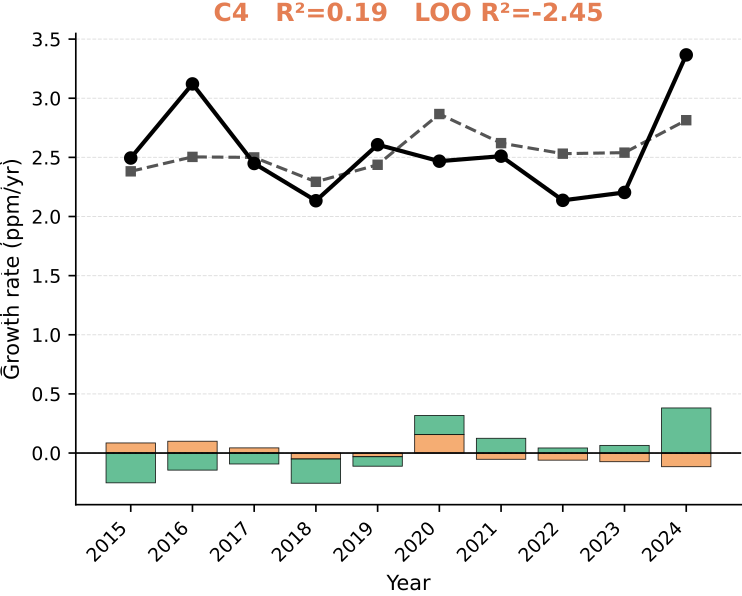}
\end{subfigure}

\begin{subfigure}[b]{.30\linewidth}
\includegraphics[width=\linewidth]{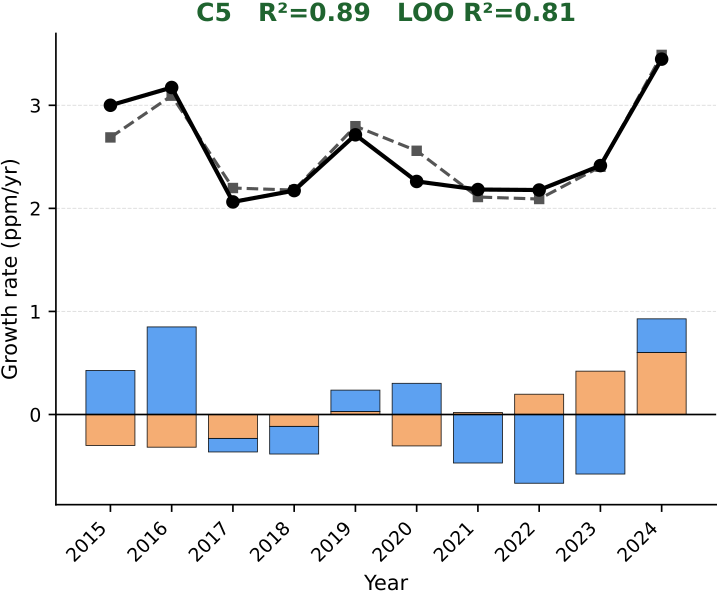}
\end{subfigure}
\begin{subfigure}[b]{.30\linewidth}
\includegraphics[width=\linewidth]{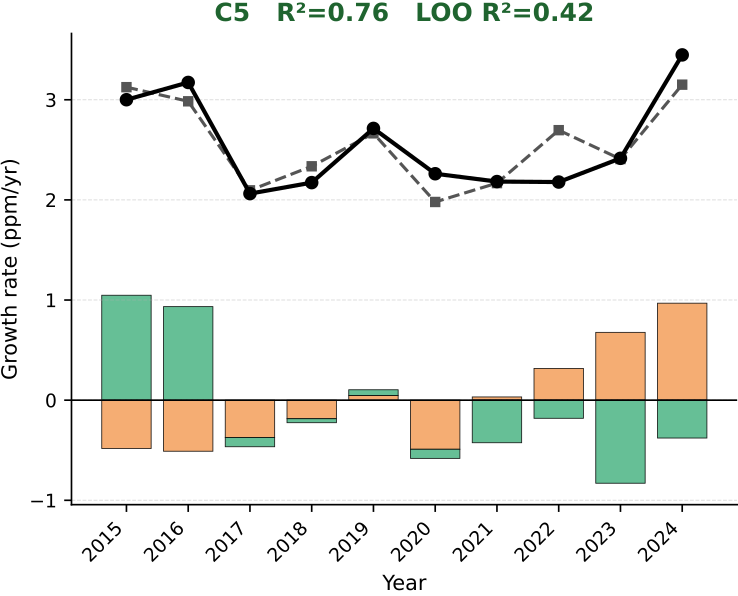}
\end{subfigure}

\caption{Annual predictor contributions to modeled CO$_2$ growth rate per persistent cluster (rows: C1–C5) for Model A (emission + SOI, left) and Model B (emission + GOSIF, right), for the years 2015–2024. Stacked bars show emission (orange) and second-predictor (blue: SOI; green: GOSIF) contributions, with the intercept excluded. The black line corresponds to the observed cluster-mean growth rate, whilst the dashed grey line corresponds to the model’s prediction. The $R^2$ and Leave-One-Out $R^2$ (LOO $R^2$) are given in panel titles. Note that the LOO $R^2$ can be negative if the model is worse than the mean.}
\label{fig:clustered-regression}
\end{figure}

Taking the clusters from the previous section as regions that share similar drivers, we investigated the effectiveness of two models that smooth over these regions, taking a more coarse grained approach to predicting CO$_2$ growth. Both models predict the CO$_2$ growth rate in terms of two drivers:
\begin{itemize}
    \item SOI Model: uses emission and SOI as drivers.
    \item GOSIF Model: uses emission and GOSIF as drivers.
\end{itemize}
Fig.~\ref{fig:clustered-regression} decomposes the annual predicted CO$_2$ growth rate for each persistent component into the contributions of human emissions $(\beta_E \cdot E_z)$ and the second climate driver — the lagged SOI in the left column and GOSIF in the right column— over the period $2015 - 2024$. The intercept is excluded from the bars so that they reflect year-to-year variability attributed to each driver rather than the mean growth-rate level. The observed and predicted CO$_2$ growth rate are shown in black and gray dashed lines respectively. 

The in-sample $R^2$ ranged from $0.43$ to $0.89$ across clusters in the SOI model, and from $0.03$ to $0.76$ across clusters in the GOSIF model, with LOO$R^2$ values between $0.01$ and $0.81$, and $ -2.45$ and $0.42$, respectively. The gap between in-sample and LOO $R^2$ was generally large for both models, suggesting some overfitting given the small number of observations used to fit the model. The SOI model outperformed the other in all of the five components, indicating that SOI is the more informative second predictor for the majority of components. In fact, the negative LOO$R^2$ in all but cluster $5$ (the biogenically-driven cluster) from the GOSIF model indicates that the GOSIF model performs worse than a constant predictor out of sample (i.e. worse than simply using the average as a predictor).

In both models, all clusters had relatively low $R^2$ scores, except for cluster $5$ (biogenically-driven cluster). This suggests that the signal associated with human emissions is more difficult to capture when averaged over a large area. Moreover, in heavily emitting components, the local CO$_2$ growth rate is influenced not just by what is emitted there, but by advection from other regions. Spatial averaging over an entire component dilutes local emission hotspots further. In contrast, in biogenic regions, the dominant flux is local and spread across the landscape, so spatial averaging actually helps by smoothing over noise.

The signal for cluster 5 is cleaner than the signal from other clusters in both the SOI and GOSIF models. Cluster 5 includes tropical and subtropical land (like the Amazon or Congo), which sits in the region most strongly influenced by ENSO effects such as the drought during El Niño suppressing uptake and directly elevating CO$_2$ growth rates. The 7-month lagged SOI is specifically calibrated to capture this response. For industrialized or high-latitude components, the ENSO effect is weaker or absent. Similarly, because the  GOSIF measures photosynthetic activity, it is a direct proxy for a process driving cluster 5, whereas for clusters that are not biogenically-driven GOSIF is a secondary signal. 

\section{Concluding remarks and Discussions}\label{sec:Conclusion}

This study provides a global, observation-based assessment of atmospheric CO$_2$ growth-rate variability and its relationship with anthropogenic emissions, biospheric activity, and large-scale climate variability. By combining globally gridded atmospheric CO$_2$ fields from CAMS reanalysis with EDGAR emissions, GOSIF-derived biospheric activity, and ENSO indicators, we move beyond administrative emission accounting frameworks and examine atmospheric carbon dynamics within a spatially continuous Earth-system perspective.

Our results show that regional atmospheric CO$_2$ growth rates arise from differing combinations of anthropogenic emissions, biospheric exchange, and climate variability, leading to distinct patterns of atmospheric response across the globe. Consequently, the detectability of emission changes is highly region dependent and cannot be inferred from observed growth rates alone. The 2020 emission reduction event illustrates how natural variability can either reveal or obscure anthropogenic signals depending on local atmospheric conditions.

By classifying regions according to their dominant controls on atmospheric CO$_2$ growth, we identify a limited number of characteristic carbon-cycle regimes that provide a physically meaningful description of global atmospheric carbon dynamics. This classification offers a foundation for improved monitoring, attribution, and prediction of regional atmospheric responses to future changes in emissions and climate.

Several limitations should be considered when interpreting these findings. First, the analysis relies on reanalysis products and associated observational constraints, which remain subject to uncertainties related to satellite retrievals, atmospheric transport modelling, and data assimilation. Second, anthropogenic emissions, biospheric activity, and ENSO represent major but not exhaustive drivers of atmospheric CO$_2$ variability. Additional factors, including regional meteorology, wildfire activity, droughts, land-use changes, ocean uptake and other modes of climate variability, may also contribute to regional atmospheric responses. Third, the statistical relationships identified here do not necessarily imply direct causality, particularly in regions where transport processes connect atmospheric signals to distant source and sink regions.

Despite these limitations, the framework developed here demonstrates the value of integrating atmospheric observations with emissions and biospheric datasets in a common spatially resolved analysis. As satellite observations continue to improve and new observing systems become available, top-down approaches will play an increasingly important role in evaluating mitigation efforts, identifying regional carbon-cycle responses, and supporting transparent climate-policy assessment.

Several avenues could further extend the framework presented in this study. First, increasing both the spatial and temporal resolution of the analysis would enable the detection of finer-scale atmospheric responses to emission changes and biospheric processes, particularly in regions with strong local heterogeneity. Second, although the CAMS reanalysis provides spatially complete and internally consistent atmospheric CO$_2$ fields, future work should directly incorporate satellite observations from missions such as OCO-2, OCO-3, GOSAT, and future CO$_2$ monitoring missions. Comparing observation-based analyses with reanalysis products will help quantify uncertainties arising from atmospheric transport models and data assimilation while strengthening confidence in top-down monitoring approaches.

In addition, extending the set of explanatory drivers beyond anthropogenic emissions, biospheric activity, and ENSO would provide a more complete description of the processes governing atmospheric CO$_2$ variability. Incorporating datasets describing wildfire emissions, ocean carbon uptake, and biospheric respiratio could improve attribution of regional atmospheric growth rates and help disentangle the influence of competing natural and anthropogenic processes. 

Finally, the regional classification introduced here opens new opportunities for predictive modelling. Rather than modelling atmospheric CO$_2$  dynamics independently at every location, future work could focus on learning the characteristic responses of each carbon-cycle regime to changes in emissions and environmental drivers. Such a cluster-based framework could improve the prediction of regional CO$_2$   evolution, identify transitions between regimes under changing climate conditions, and provide a more interpretable and computationally efficient representation of the global carbon cycle. However, our results indicate that these clusters must be further refined to capture unique regional signals, as current spatial averaging tends to smooth over distinct patterns, rendering them indistinguishable within the margin of error.  Integrating the additional drivers mentioned above would likely provide the necessary granularity to make these clusters more unique and distinguishable.

\section{Data availability}
The implementation code and data for this study can be found on GitHub at \url{https://github.com/ghg-ai-lab-complexity72-dev/complexity72_2026}.

\section{Acknowledgements}
This work is the output of the workshop Complexity72h by Complexity Next Gen, held at Northeastern University London, London, UK, 22-26 June 2026. www.complexitynextgen.org/complexity72h/.

We thank Saba Minhaj for contributing to the code development for applying the DLM code to the CAMS gridded dataset to produce the growth rate dataset.

\bibliographystyle{unsrtnat}
\bibliography{references} 
\clearpage


\section*{Supplementary material}

## 3 Methods for determining the optimal cluster number

We use two complementary methods to determine the optimal cluster number $k^*$ for each year. Both methods yield consistent results.

The **elbow method** Thorndike [1953] picks an optimal number of clusters $k$ by checking how the intra-cluster variance $v(k) = \sum_{i=1}^{k} \sum_{x \in C_i} ||x - \mu_i||^2$ decreases with $k$, and identifying the critical value of $k$ where adding more clusters stops paying off proportionally. A convenient way to locate the critical value is to compute, for each $k$, the perpendicular distance $d(k)$ from $(k, v(k))$ to the straight line connecting $(k_{\min}, v(k_{\min}))$ and $(k_{\max}, v(k_{\max}))$. The optimal $k$ maximizes this distance: $k^* = \arg\max_k d(k)$.

The **silhouette method** Rousseeuw [1987] compares two distances for each data point $x_i$ in cluster $C_i$, $i = 1, ..., k$:

(i) intra-cluster distance as the mean distance from $x_i$ to al other points in its own cluster: $a(x_i) = \frac{1}{|C_i|-1} \sum_{x_j \in C_i, j \neq i} ||x_i - x_j||$ and

(ii) nearest-cluster distance as the mean distance from $x_i$ to all points in the closest other cluster minimized over all clusters: $b(x_i) = \min_{k \neq i} \frac{1}{|C_k|} \sum_{x_j \in C_k} ||x_i - x_j||$.

The silhouette coefficient of the point $x_i$ is defined as $s(x_i) := \frac{b(x_i) - a(x_i)}{\max\{a(x_i), b(x_i)\}} \in [-1, 1]$, and the silhouette score for a given $k$ is the average over all $N$ points: $S(k) = \frac{1}{N} \sum_{i=1}^{N} s(x_i)$. The optimal $k$ is given by $k^* = \arg\max_k S(k)$.

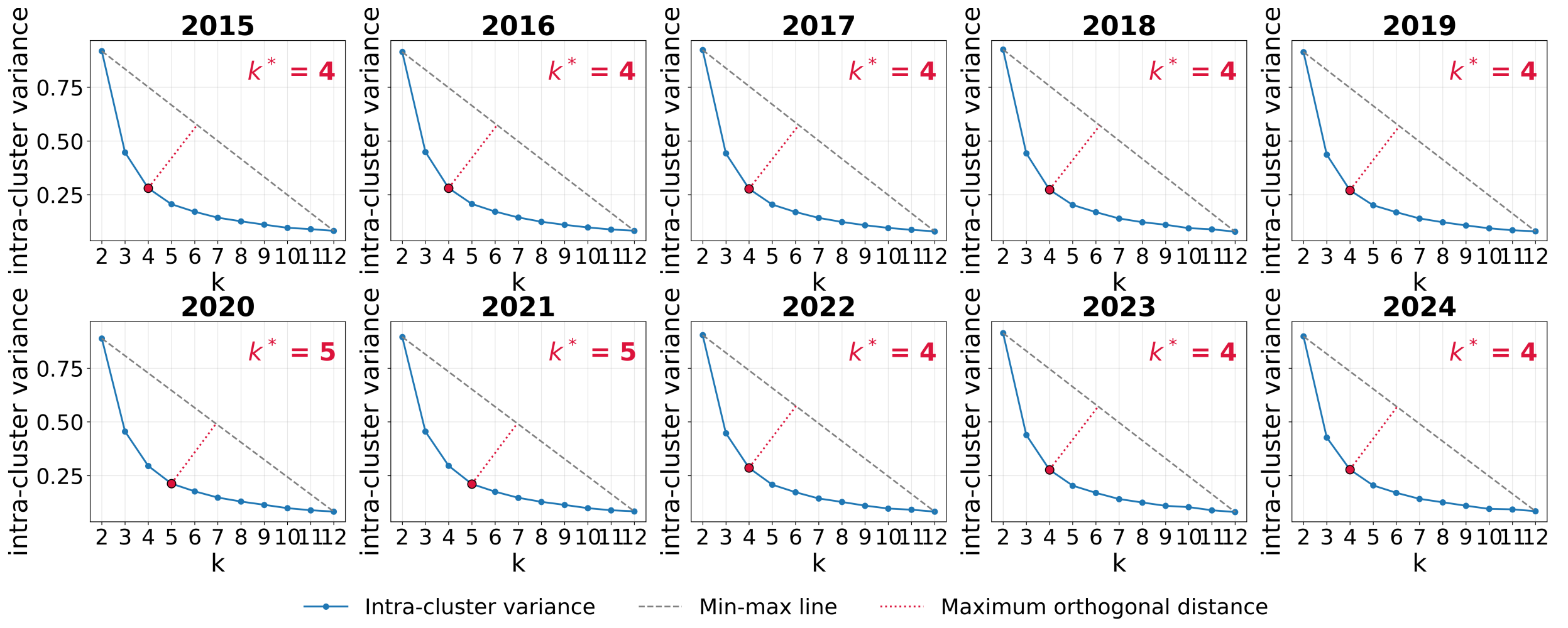


Figure 7: Elbow method: Intra-cluster variance as a function of $k$ for each year. The optimal cluster size $k^*$ is highlighted in red.

The dataset consists of annual global gridded observations at $1° \times 1°$ resolution (180 latitude×360 longitude) for the period 2015–2024. Each grid cell is characterized by a tuple (DSe, DSb). Ocean grid cells are excluded from the analysis. For each year, we perform the K-means with $k \in \{2, 3, ..., 12\}$ and determine the optimal cluster number $k^*$ using both methods.

Fig. 7 shows the result of the elbow method for each year, where the optimal $k$ is highlighted in red: apart from the years 2020 and 2021, where $k^* = 5$, for the rest years we have $k^* = 4$.

In Fig. 8, we perform the silhouette method for each year, where the optimal $k$ is highlighted in red. We have $k^* = 6$ for the years 2020 and 2021, $k^* = 3$ for year 2024, and $k^* = 4$ for the rest of the years.

The resulting optimal $k^*$ oscillates between 4 and 6 across the ten years.

## 4 Acknowledgements

This work is the output of the workshop Complexity72h by Complexity Next Gen, held at Northeastern University London, London, UK, 22-26 June 2026. www.complexitynextgen.org/complexity72h/.

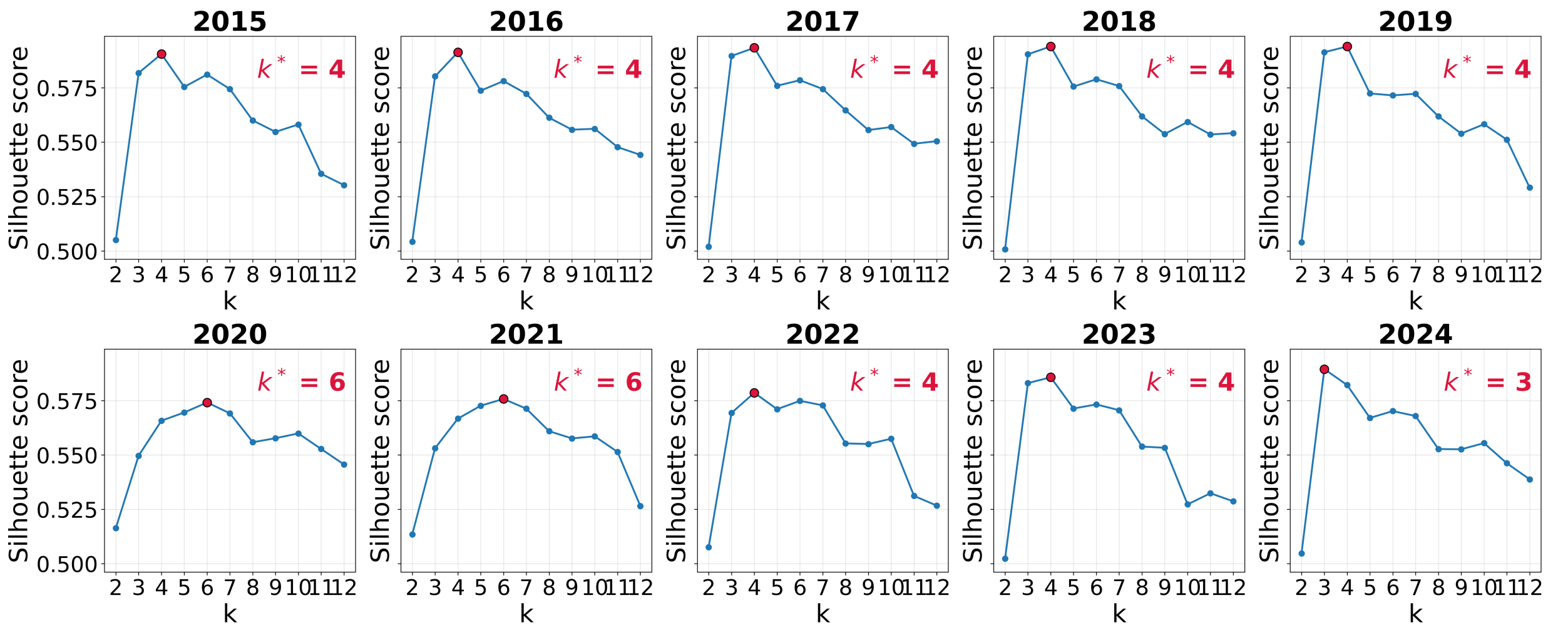


Figure 8: Silhouette method: Silhouette score as a function of $k$ for each year. The optimal cluster size $k^*$ is highlighted in red.



\end{document}